\documentclass{llncs}

\usepackage{url,wrapfig}

\usepackage{pgf}
\usepackage{tikz}
\usetikzlibrary{arrows,automata}
\usepackage[latin1]{inputenc}

\newcommand{\code}[1]{{\small {\ensuremath{\tt #1}}}}

\newcommand{\reff}[1]{(\ref{#1})}

\newcommand{\while}{\code{while}}

\newcommand{\tab}{\quad}

\newcommand{\transrel}{\rho}

\title{{Preliminary Notes on Termination and Non-Termination Reasoning}}
\author{Ton Chanh Le}
\institute{Department of Computer Science,\\ National University of Singapore\\
\texttt{chanhle@comp.nus.edu.sg}}

\begin{document}
  \maketitle
  
  \begin{abstract}
In this preliminary note, we will illustrate our ideas on automated
mechanisms for termination and non-termination reasoning. 
\end{abstract}

  \section{Termination and Non-Termination Reasoning with Logical Abduction}
  \subsection{Introduction}
Program termination and non-termination reasoning have gained
an enormous interest over the last decade. For termination reasoning,
beside the traditional approach with ranking functions, there are
various new approaches are proposed, such as the size-change principle
for functional programs \cite{Lee:POPL01}, the polyranking principle for multi-path
programs \cite{Bradley:ICALP05} and the disjunctive wellfoundedness principle 
for imperative programs \cite{Cook:PLDI06}.
For non-termination reasoning, Gupta {\em et al.} \cite{Gupta:POPL08}
propose a constraint-based approach to find the recurrent set of a loop, 
which is the initial configuration for infinite executions of this loop.

However, in these representative works, termination and non-termination analyses 
are usually considered as separate mechanisms. In other words, there is
no cooperation between termination and non-termination analyses in these
works. Inspired from the recent success of the cooperation between safety and termination
provers in finding supporting invariants for the validity of termination arguments
\cite{Brockschmidt:CAV13,Larraz:FMCAD13,Bradley:CAV05}, we strongly believe that 
the cooperation between termination and non-termination provers will gain the same
success as the termination analysis can benefit from the non-termination analysis 
by excluding non-terminating behaviors of programs and vice-versa.
For example, the idea of excluding non-terminating program behaviors is important
for termination analyzers which is based on counterexample to termination to
construct termination arguments like ARMC \cite{Podelski:PADL07:ARMC}. 
These tools might not terminate when dealing with non-terminating loops
as the number of the generated counterexamples is infinite.

There are few past works that take advantage of non-termination analysis in
reasoning program termination. In \cite{Harris:SAS10}, the supporting invariants
for termination arguments are synthesized as the complement of unreachable 
non-terminating conditions by relying on an existing non-termination prover
\cite{Gupta:POPL08} and a safety prover. In \cite{Bozga:TACAS12}, the authors
determine in which domains the weakest precondition for non-termination is 
decidable, so does the precondition for termination. \cite{Ganty:CAV13} is another 
related approach, which divides a transition relation into two sub-relations
whose termination is proved and whose behaviors are still unknown by a fixpoint 
computation. A program is proved to be terminating for all inputs if the set
of unknown behaviors is empty. Otherwise, a sufficient condition for termination 
is returned.

As opposed to these approaches, our aim is to solve not only the conditional 
termination problem \cite{Cook:CAV08} but also the conditional non-termination 
problem so that we can construct a more complete picture about 
the terminating and non-terminating behaviors of a program.
In general, our proposed mechanism will incrementally partition unknown program 
behaviors into terminating, non-terminating and/or unknown sub-behaviors.
The analysis can be facilitated by (safety) preconditions of the program
provided by users or by automated safety provers. Otherwise, the analysis
will start with the trivial precondition \code{true}.

Specifically, we initially prove the program non-termination by showing
that its exit points are unreachable under the given preconditions.  
If not, then the program might terminate. 
Next, a ranking function synthesizer, such as \cite{Podelski:VMCAI04}, 
will be invoked to find a termination measure for the program. 
As the possibility of non-termination is excluded,
when the complete method for linear ranking function synthesis \cite{Podelski:VMCAI04} 
fails, there are two remaining possibilities that could happen: 
(i) the program terminates but a linear ranking function does not exist or 
(ii) a supporting invariant for termination proving is missing. 
For both possibilities, we then perform a case analysis on the program preconditions 
by an auxiliary condition, called {\em potential non-terminating condition} and its negation.

Intuitively, that condition ensures that the program does not terminate 
(or the loop condition is not violated) after the current execution and it
can be inferred by {\em logical abduction}. Here we do not seek an actual
non-terminating precondition like \cite{Harris:SAS10} as it is still 
possible that the program terminates (without a linear ranking function).
If the inferred condition is an actual precondition for non-termination, 
it will be proved in the next step of the analysis. Otherwise, it will be
continuously refined in further steps. In both cases, the negation of the
potential non-terminating condition also facilitate the termination reasoning
as it strengthens the precondition in which the program's exit points are 
reachable.

\subsection{Examples}

Let consider two resembling examples with different (non-) termination behaviors 
and their corresponding loop transition relations in Fig. \ref{fig:examples}. 
To analyze the termination and non-termination of these loops, we are only interested 
in the non-trivial cases, {\em i.e.}, when the loops' conditions are satisfied 
and these loops thus execute at least one time.

\begin{figure}[thb]
\begin{small}
\begin{center}
\begin{tabular}[t]{|c|c|}
\hline
\begin{array}[t]{c}
\code{\while ~ (x{\geq}0) ~\{}
\code{~ x = x{-}y;}
\code{~ y = y{+}1; ~\}}
\\\\
\transrel_a \equiv x{\geq}0 \wedge x'{=}x{-}y \wedge y'{=}y{+}1
\end{array}
&
\begin{array}[t]{c}
\code{\while ~ (x{\geq}0) ~\{}
\code{~ x = x{-}y;}
\code{~ y = y{-}1; ~\}}
\\\\
\transrel_b \equiv x{\geq}0 \wedge x'{=}x{-}y \wedge y'{=}y{-}1
\end{array}
\\
(a) & (b)\\
\hline 
\end{tabular}
\end{center}
\end{small}
\vspace{-4mm}
\caption{Examples on numerical programs with different termination behaviors. 
Example (a) is terminating for any input while example (b) is always non-terminating
for the input $x{\geq}0 \wedge y{\leq}0$.}\label{fig:examples}
\end{figure}

For those examples, we initially prove their non-termination by showing that
their exit points are {\em unreachable}. This can be done by an unsatisfiability 
check on the formula $\transrel \wedge \neg \psi[X'/X]$, 
where $\transrel$ and $\psi$ are the transition
relation and the condition of the loop, respectively. For instance, the exit point
of the loop in Fig. \ref{fig:examples}(a) might be reachable from the loop body
as the following formula
\begin{equation}\label{reach-loop-a}
\transrel_a \wedge \neg(x' \geq 0) \equiv x \geq 0 \wedge x'=x+y \wedge y'=y-1 \wedge x' < 0
\end{equation}
is satisfiable, which means that the program is possibly terminating. 

We then check whether the loop terminates immediately after the current execution 
by the validity of the entailment 
\begin{equation}\label{term-loop-a}
\transrel_a \vdash \neg(x' \geq 0) \equiv x \geq 0 \wedge x'=x+y \wedge y'=y-1 \vdash x' < 0
\end{equation}
The entailment \reff{term-loop-a} is invalid or the formula
$\transrel_a \wedge x'{\geq}0$ is satisfiable. This result shows that with the 
precondition $x{\geq}0$, satisfying the loop condition, the loop might execute in more than one steps.
We now know that we need a ranking function synthesis tool, such as \cite{Podelski:VMCAI04}, 
to find a witness for the program termination, if any. With two above simple checks
on \reff{reach-loop-a} and \reff{term-loop-a}, we can avoid to invoke the synthesis tool
too early on trivial cases of termination or when the considered loop is non-terminating.

However, we receive a failure result from the synthesis tool. Consequently,
there is not a linear ranking function for that loop and/or a supporting
invariant, which helps to separate terminating and non-terminating execution of the loop, 
is missing for termination proving on that program. 
To find out the missing invariant, we firstly infer a {\em potential non-terminating condition},
which ensures that the loop condition is not violated after the current execution.
As opposed to the condition inferred from the finding of a potential ranking function 
\cite{Cook:CAV08,Ganty:CAV13}, this condition can be logically inferred from
the entailment $\transrel \vdash \psi[X'/X]$. Thus, we can handle loop conditions with
a more general structure ({\em e.g.}, disjunctions).

A trivial solution for such potential non-terminating condition is $\psi[X'/X]$, where
$\psi$ is the loop condition. Thus, the negation of this condition is a sufficient condition
for program termination, under which the loop terminates after the current execution.
However, we need a stronger condition which can be
obtained via {\em logical abduction} \cite{Peirce:Book74}. An abductive constraint $\phi$ for
the validity of the entailment $\Gamma \vdash \psi$ is usually inferred by the quantifier elimination 
on the universal formula $\forall V.(\neg \Gamma \vee \psi)$ where the set of variables $V$
can be determined from a minimum satisfying assignment (MSA) of $\Gamma \vdash \psi$  as in 
\cite{Dillig:CAV13} or defined by users. Another approach which is suitable for abductive inference 
is constraint solving. In this approach, the unknown constraint $\phi$ would be inferred by Farkas' 
lemma and a non-linear solver.
In contrast, we have designed a lightweight abductive inference for linear arithmetic
with a balance between the number of variables and the generality of the inferred constraints.
Our approach does not require an expensive quantifier elimination with a calculation of MSA or
non-linear constraint solver. 

Back to the example \ref{fig:examples}(a), the inferred potential non-terminating condition
is $y \leq 0$ as $\transrel_a \wedge y{\leq}0 \vdash x'{\geq}0$. 
Then we perform a case analysis on this condition by
partitioning the transition relation $\transrel_a$ into 
$\transrel_a^1 \equiv \transrel \wedge y{\leq}0$ and 
$\transrel_a^2 \equiv \transrel \wedge y{>}0$. 
As opposed to our approach, \cite{Ganty:CAV13,Larraz:FMCAD13} syntacticly 
partition the transition relation based on the condition for the decrease
of a potential ranking function or the boundedness of a quasi-ranking function, 
respectively.

Applying the same procedure
for $\transrel_a^1$ and $\transrel_a^2$, we can prove that the execution corresponding
to the relation $\transrel_a^2$ (when $x{\geq}0 \wedge y{>}0$) always terminates.
The termination proof is successful because the auxiliary condition $y > 0$ is also
proved to be a supporting invariant.  
The execution corresponding to the relation $\transrel_a^1$ (when $x{\geq}0 \wedge y{\leq}0$)
also terminates and eventually reaches the terminating case $x{\geq}0 \wedge y{>}0$.
Thus, the program always terminate for any input.

For the example \ref{fig:examples}(b), we can similarly infer the same potential 
non-terminating condition $y \leq 0$. This condition is also the actual non-terminating
condition of the loop. For the other case when $x{\geq}0 \wedge y{>}0$, we mark it as possibly
terminating case as the execution from it can reach either the base case $x{\geq}0$ or the non-
terminating case $x{\geq}0 \wedge y{\leq}0$ according to the initial reachability checks.

\begin{wrapfigure}{L}{30mm}
\vspace{-1cm}
\begin{center}
\begin{small}
\begin{array}[t]{|l|}
\hline
\code{y = f();}\\
\code{\while ~ (x \geq 0) ~\{}\\
\code{\tab x = x-y;}\\
\code{\tab y = y+1; ~\}}\\
\hline
\end{array}
\end{small}
\caption{A modified version of an example from \cite{Brockschmidt:CAV13}
with a complicated function $f$.}\label{fig:modular}
\end{center}
\vspace{-1.2cm}
\end{wrapfigure}

Another advantage of this approach is that the (non-) termination of each loop is analyzed in 
a modular fashion as the supporting invariants for termination proof are discovered from 
the loop itself rather than from its initial contexts, which might be too complicated to 
analyze. Let consider a modified version of an example from \cite{Brockschmidt:CAV13}
in Fig. \ref{fig:modular}, in which the initial value of $y$ is determined by a complicated function
$f$. Consequently, the supporting invariant $y>0$ for the ranking function $x$ is hard
to discover due to the complexity of the analysis on $f$. For this example, we analyze
the while loop in isolation and show that the loop terminates for all input as with the
example in Fig. \ref{fig:examples}(a). Thus, we can conclude that the whole program
terminates with an assumption that the function $f$ also terminates.

%
%

  
  \bibliography{ref}
  \bibliographystyle{abbrv} 
\end{document}